\documentclass[twocolumn, showpacs,footinbib]{revtex4}
\usepackage[T1]{fontenc}
\usepackage{ae}
\usepackage{aecompl}
\usepackage[latin1]{inputenc}
\usepackage{graphicx}
\usepackage{amssymb}
\usepackage[normalem]%
{ulem}  
\usepackage[dvips]{color}

\newcommand{\beq}{\begin{equation}}
\newcommand{\eeq}{\end{equation}}
\newcommand{\bqa}{\begin{eqnarray}}
\newcommand{\eqa}{\end{eqnarray}}
\newcommand{\be}{\begin{equation}}
\newcommand{\ee}{\end{equation}}
\newcommand{\bea}{\begin{eqnarray}}
\newcommand{\eea}{\end{eqnarray}}
\newcommand{\nn}{\nonumber}
\newcommand{\0}{\over }


\begin{document}

\title{Non-Abelian plasma instabilities: SU(3) vs. SU(2)}

\preprint{TUW-10-17}

\author{Andreas Ipp}

\affiliation{Institut f\"ur Theoretische Physik, Technische Universit\"at Wien,
Wiedner Hauptstra\ss e 8-10, A-1040 Vienna, Austria}

\author{Anton Rebhan}

\affiliation{Institut f\"ur Theoretische Physik, Technische Universit\"at Wien,
Wiedner Hauptstra\ss e 8-10, A-1040 Vienna, Austria}

\author{Michael Strickland}

\affiliation{
Department of Physics, Gettysburg College, Gettysburg, PA 17325, USA
}

\begin{abstract}
We present the first 3+1 dimensional simulations of non-Abelian 
plasma instabilities in gauge-covariant Boltzmann-Vlasov equations
for the QCD gauge group SU(3) as well as for SU(4) and SU(5). 
The real-time evolution of instabilities
for a plasma with stationary momentum-space anisotropy is studied
using a hard-loop effective theory that is discretized in the velocities
of hard particles. 
We find that the numerically less expensive calculations using
the group SU(2) essentially reproduce the nonperturbative dynamics of
non-Abelian plasma instabilities with higher rank gauge groups
provided the mass parameters of the
corresponding hard-loop effective theories are the same. In particular
we find very similar spectra for the turbulent cascade that forms
in the strong-field regime, which is associated with an approximately 
linear growth
of energy in collective fields. 
The magnitude of the linear
growth however turns out to increase with the number of colors.
\end{abstract}

\pacs{11.15.Bt, 04.25.Nx, 11.10.Wx, 12.38.Mh}

\maketitle

\section{Introduction}

Non-Abelian plasma instabilities are, parametrically, the dominant collective
phenomenon in a weakly coupled quark-gluon plasma, and have been discussed
as a possible explanation for the extremely fast 
isotropization that is suggested by the success of hydrodynamical models of
relativistic heavy-ion collisions \cite{Heinz:2004pj}. 
Such plasma instabilities
are generalizations of the so-called Weibel or filamentary instabilities
in ordinary electromagnetic plasmas \cite{Weibel:1959zz}. They
are present already in collisionless plasmas with any amount of momentum
space anisotropy \cite{Romatschke:2003ms,Romatschke:2004jh}, and, indeed,
they have been found to play an important role in the fast isotropization
of electromagnetic plasmas \cite{Califano:2001}. Their non-Abelian
versions have been proposed to be of relevance for the quark-gluon
plasma early on by Mr\'owczy\'nski and others \cite{Mrowczynski:1988dz,Mrowczynski:1993qm,Mrowczynski:2000ed,Randrup:2003cw,
Pokrovsky:1988bm,Pokrovsky:1990sz,Pokrovsky:1990uh,Pavlenko:1991ih,
Mrowczynski:1994xv,Mrowczynski:1996vh},
and specifically as explanation for the fast apparent
thermalization by Arnold et al.\ 
\cite{Arnold:2003rq,Arnold:2004ti}.

Numerically, these instabilities have been studied in a discretized version
of the hard-loop approximation \cite{Rebhan:2004ur,Arnold:2005vb,Rebhan:2005re}, which corresponds to gauge covariant Boltzmann-Vlasov equations describing
the dynamics of (soft) collective fields in a weakly coupled plasma of
hard particles. The growth rate of plasma instabilities is parametrically
of the same order as plasma frequencies and screening masses, and because
of their exponential behavior the plasma instabilities dominate
the collective dynamics and inevitably lead to nonperturbatively large
collective fields. Eventually they will give rise to
substantial backreactions on the momentum distribution of the hard
particles, causing a breakdown of the hard-loop approximation
\cite{Dumitru:2007rp,Dumitru:2006pz,Dumitru:2005gp} 
coincident with the actual isotropization process.
The hard-loop approximation allows to study the early stage of
this assumed scenario and thus its basis. 

In the earliest stages of heavy-ion collisions it is in fact
important to take into account the expansion of the plasma, which
modifies the growth from exponential in time to exponential in
the square root of (proper) time, as indeed found in numerical
simulations within the color glass condensate scheme 
\cite{Romatschke:2005pm,Romatschke:2006nk} as well as in
a generalization of discretized hard loop simulations 
\cite{Romatschke:2006wg,Rebhan:2008uj}. In those simulations,
an uncomfortable delay of the onset of growth has been observed,
which however has recently been shown to largely disappear
upon consideration of more general initial conditions
\cite{Rebhan:2009ku}.
While the density and lifetime of the plasma estimated for
heavy-ion collisions at RHIC may be too low to give an important
role to nonabelian plasma instabilities there, the higher values
expected for LHC heavy-ion collisions may be sufficient
for nonabelian plasma instabilities to become the dominant
phenomenon in a less strongly coupled environment.

By means of real-time lattice simulations for stationary anisotropic plasmas, 
it has however
been found that in contrast to effectively 1+1 dimensional situations
where only the most unstable modes are considered,
the exponential growth of non-Abelian plasma instabilities 
is limited in 3+1 dimensions
by non-Abelian self-interactions. 
At a certain magnitude of the nonabelian fields, which depends
on the degree of anisotropy \cite{Bodeker:2007fw,Arnold:2007cg},
the exponential growth ceases and turns into a linear growth of
the energy densities of the soft fields. In that regime, a turbulent
cascade of energy is observed, with a
quasi-steady-state power-law distribution 
$f_k \propto k^{-\nu}$ and spectral index $\nu \approx 2$,
which transports the energy fed into low-lying modes by Weibel instabilities
to stable higher-momentum (plasmon) modes through
non-Abelian self-interactions of gluon fields \cite{Arnold:2005ef}.
This cascade forms at momentum scales that are parametrically separated
from those of the hard particles, making it possible to study
this phenomenon self-consistently within the hard-loop approximation.
On the other hand, classical-statistical simulations in
SU(2) gauge theory \cite{Berges:2007re,Berges:2008mr}, where there
is no such separation of scales, have
reported a late-time behavior indicative of a lower index
of $\nu=4/3$ (as in turbulence
with constant transport of particle number)
while showing an early behavior qualitatively similar
to those of Chromo-Weibel plasma instabilities. The index $\nu=2$ found
in the hard-loop simulations was argued \cite{Arnold:2005qs}
to fit to an energy cascade carried by particles in the cascade
scattering off of nonperturbatively large
background fields. However, this is not consistent with a steady-state particle cascade
\cite{Mueller:private}
(particle number of cascade particles is unchanged in
the assumed process) so that either other processes are equally
important or the resulting cascade is not steady-state.
(See also Refs.\ \cite{Bodeker:2005nv,Dumitru:2005gp,Dumitru:2006pz,Mueller:2005un,Mueller:2005hj,Asakawa:2006tc,Mueller:2006up,Majumder:2006wi,Carrington:2010sz} for discussions of instabilities and turbulence in QCD.)

In this paper we confirm 
the results \cite{Arnold:2005vb,Rebhan:2005re,Arnold:2005ef,Bodeker:2007fw,Arnold:2007cg} obtained for stationary
anisotropic plasmas in the hard-loop framework for gauge group SU(2),
and consider, for the first time, 3+1-dimensional hard-loop
simulations for the QCD gauge group SU(3), as well as for SU(4)
and SU(5), in order to quantify the dependence on the number of colors.
So far, in the hard-loop effective theory,
SU(3) calculations have only been performed for
effectively 1+1 dimensional situations 
\cite{Rebhan:2005re}. (In classical-statistical 3+1-dimensional
lattice gauge theory the gauge group SU(3) has been studied in 
Ref.~\cite{Berges:2008zt}.)

\section{Setup}

In order to study nonabelian plasma instabilities in a weakly
coupled quark-gluon plasma, which may for the first time become
physically relevant at the higher temperatures and densities
reached in heavy-ion collisions at the LHC, we consider the
extreme limit of an ultrarelativistic collisionless plasma,
where the main dynamics takes place at scales parametrically
soft compared to the
hard scale $|\mathbf p|=p^0$ of the plasma constituents.
In an isotropic plasma, the scale $g|\mathbf p|$, where $g$
is the gauge coupling, determines
the scale of the Debye screening mass, of the plasma frequency,
and of Landau damping, but in an anisotropic plasma, the
dominant collective phenomenon at this scale, which is larger
than the scale of collisions, turns out to be plasma instabilities.

The effective field theory relevant for the collective phenomena
at this largest of the soft scales is given by gauge-covariant
collisionless Boltzmann-Vlasov equations 
\cite{Blaizot:2001nr}
or ``hard-loop'' effective theory as long as 
the soft gauge fields obey $A_\mu \ll |\mathbf p|/g$
so that a backreaction on the hard particles can still be
ignored. This allows one to study in detail the first stage
of plasma instabilities, which, as briefly reviewed in the introduction,
is highly nontrivial in nonabelian gauge theories.

The corresponding effective action, which is nonlocal and nonlinear
\cite{Pisarski:1997cp,Mrowczynski:2004kv}, can be made local
at the expense of introducing auxiliary fields \cite{Blaizot:1993be}
in the adjoint
representation, $W_\beta(x;\mathbf v)$, for each spatial
unit vector appearing in the velocity $v^\mu=p^\mu/|\mathbf p|=(1,\mathbf v)$
of a hard (ultrarelativistic) particle with momentum $p^\mu$.
The $W$ fields
encode the fluctuations of the distribution function of
colored hard particles. In terms of these, the induced current $j$
in the nonabelian Maxwell equations
\begin{equation}
\label{Feq} D_\mu(A) F^{\mu\nu}=j^\nu,
\end{equation}
is given by 
\begin{equation}
\label{Jind0} j^\mu[A] = -g^2 \int {d^3p\over(2\pi)^3} 
{1\over2|\mathbf p|} \,p^\mu\, {\partial f(\mathbf p) \over \partial p^\beta} W^\beta(x;\mathbf v) ,
\end{equation}
and the nonabelian Boltzmann-Vlasov equation, in which the
scale of the ultrarelativistic hard particles drops out, reduces to
\begin{equation}
\label{Weq} [v\cdot D(A)]W_\beta(x;\mathbf v) = F_{\beta\gamma}(A) v^\gamma ,
\end{equation}
with $D_\mu=\partial_\mu-ig[A_\mu,\cdot]$. (Our metric convention is 
$(+---)$.)

We shall only consider a background distribution function $f(\mathbf p)$
of hard particles with one direction of anisotropy, obtained by
deforming an isotropic distribution according to
\be
f(\mathbf p) \propto f_{\rm iso}(\mathbf p^2+\xi p_z^2)
\ee
which leads to
\bea\label{jmuab}
j^\mu(x)&=&\int{d\Omega_{\mathbf v} \over 4\pi}\left[
a(\mathbf v)W^0(x;\mathbf v)+b(\mathbf v)W^z(x;\mathbf v)
\right]v^\mu\nn\\
&\equiv&
\int{d\Omega_{\mathbf v} \over 4\pi}\mathcal W(x;{\mathbf v})v^\mu,
\eea
and
\be\label{avbvxi}
a({\mathbf v})={m^2\0(1+\xi v_z^2)^2},\qquad
b({\mathbf v})=\xi v_z a({\mathbf v}),
\ee
with the mass parameter $m$ proportional to $g$ and the scale of the momenta of hard particles.

In the numerical treatment we discretize the 3-dimensional
configuration space by cubic lattices, and the 
unit sphere by a set of unit vectors $\mathbf v$ pointing
to 
\bea\label{disco}
&&z_i=-1+(2i-1)/N_z,\quad i=1\ldots N_z,\nn\\
&&\varphi_j=2\pi j/N_\varphi,\quad j=1\ldots N_\varphi
\eea
in cylindrical coordinates.
At each lattice site we thus introduce $\mathcal N_{\mathcal W}=\mathcal
N_z\times \mathcal N_\varphi$
variables $\mathcal W_{\mathbf v}(x)$ whose dynamics is
approximated by \cite{Rebhan:2004ur,Rebhan:2005re}
\bea\label{DHLW}
&&[v\cdot D(A)]\mathcal W_{\mathbf v}=(a_{\mathbf v} F^{0\mu}+b_{\mathbf v}
F^{z\mu})v_\mu\\
\label{DHLF}
&&D_\mu(A) F^{\mu\nu}=j^\nu={1\0\mathcal N}\sum\limits_{\mathbf v} v^\nu 
\mathcal W_{\mathbf v}.
\eea
A different possibility of discretization was used in 
Ref.~\cite{Arnold:2005vb},
where the auxiliary fields $W^\mu(x;\mathbf v)$ are expanded in spherical
harmonics $Y_{lm}(\mathbf v)$ and truncated at some $l_{max}$.

In the following we shall consider a momentum space distribution
of hard modes which is oblate, by choosing $\xi=10$  
as before in Refs.~\cite{Rebhan:2004ur,Rebhan:2005re}.
In this situation, a prolate region in momentum space at
soft scales, pinched at the origin and involving 
longitudinal momentum $0<|k_z|<\mu=1.072 m$, contains
unstable modes. The mode with largest growth rate $\gamma_*=0.12 m$
is found at $\mathbf k_\perp=0$ and $k_z=k_*=0.398 m$. The mass
parameter $m$ defined above is related to the asymptotic mass
of propagating transverse gluons by $m_\infty=0.447 m$.

The details of the lattice discretization of the above
nonabelian gauge-covariant Boltzmann-Vlasov equations
can be found in the appendix of Ref.~\cite{Rebhan:2005re}.
The only place where those formulae need to be modified to
allow for general $N_c$ is
in the expression for the energy density.
For arbitrary $N_c$, the chromo-magnetic part of the 
energy density in Eq.~(B18)
of Ref.~\cite{Rebhan:2005re} reads
\be
\mathcal{E}_B  =  \frac{2N_c}{g^{2}a^{4}}\sum_{\square}\left(1-\frac{1}{N_c}{\rm tr}\,U_{\square}(t,x)\right).
\ee

In the numerical simulations of the above equations we have
used parallelization on computer clusters, splitting
up the simulation domain so as to fit into the main memories
of the individual nodes, and considered total spatial lattices
ranging from $32^3$ to $128^3$ and numbers
of $\mathcal W$ fields 
$\mathcal N_{\mathcal W}=\mathcal N_z\times \mathcal N_\varphi=
20\times 16=320$ and higher
at each site, in addition to
the $N_g=N_c^2-1$ gluon fields (link variables) and their conjugate momenta.

The physical size of the lattice is determined by the parameter $m^2$
appearing as prefactor in the induced current, Eq.~(\ref{jmuab}). In terms
of the asymptotic thermal mass of gluons, the lattice spacings that
we have considered vary from $a m_\infty=
0.5656$ 
to $0.1414
$.
For the specific anisotropy that we are considering, this corresponds
to
$a/(\lambda_*/2)=0.16 \ldots 0.04$
in terms of the wave length of maximal growth $\lambda_*$,
or, with respect to the minimal wave length $\lambda_\mu$
for which unstable modes
exist: $a/(\lambda_{\mu}/2)=0.432 \ldots 0.108$.

In order to provide seed fields, we have initialized with
vanishing gauge fields and
\beq\label{3dinitcond}
\left<{\mathcal W}^a_{\bf v}(0,x) {\mathcal W}^b_{\bf v}(0,y)\right>=
\delta^{ab}\delta^3_{x,y}\sigma^2 
a^3,
\eeq
where $\delta^3_{x,y}$ denotes 3 Kronecker deltas, subtracting
$\sum_{\bf v} {\mathcal W}^a/{\mathcal N_{\mathcal W}}^2$ from each ${\mathcal
W^a}$  in order to obey
Gauss's law.

\section{Numerical results}

\subsection{Growth of energy densities in chromo-electromagnetic fields}

In the following we shall concentrate on the case of small initial
fields (small $\sigma$) so that there is first a phase of essentially
Abelian evolution with exponential growth due to the Weibel instability,
which ends when the gauge
fields have grown to nonperturbatively large amplitudes
so that non-Abelian self interactions become important.
We keep fixed the anisotropy parameter $\xi=10$ in all of the
following comparisons.

\begin{figure}
\begin{center}.\includegraphics[%
  clip,
  scale=0.65]{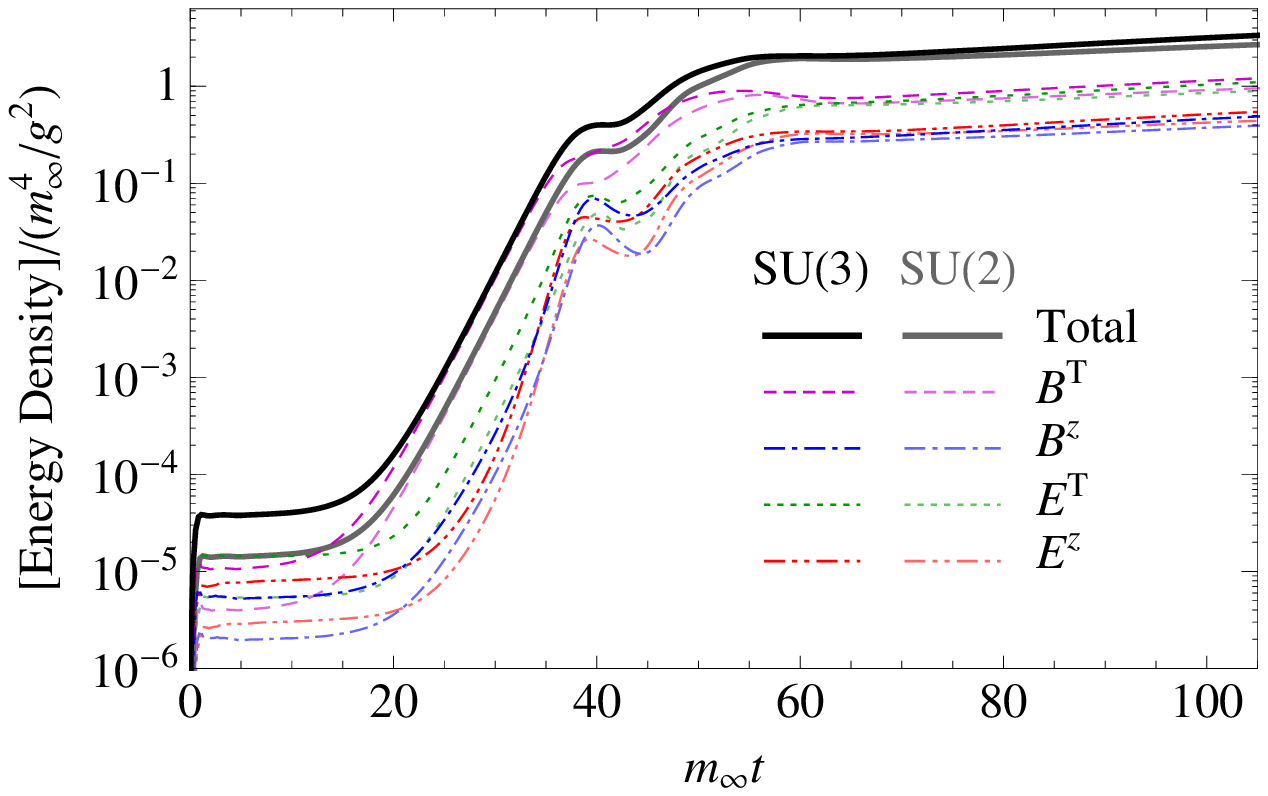} .\includegraphics[%
  clip,
  scale=0.65]{
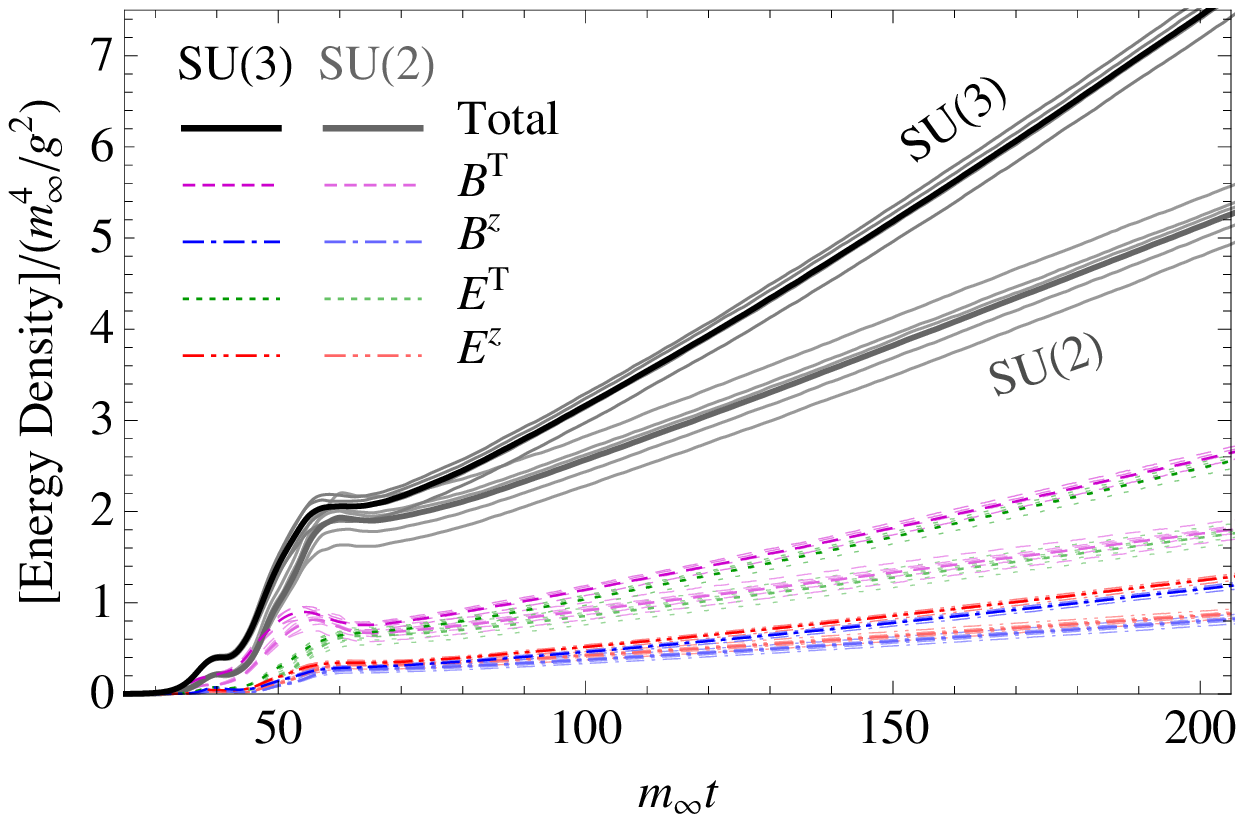}\end{center}
\caption{(Color online.) Comparison of average energy densities $\mathcal{E}$ for SU(2) (light)
and SU(3) (dark) on logarithmic (upper panel) and linear (lower panel) scale in
3+1 dimensional simulations for anisotropy parameter $\xi=10$ on
a $64^{3}$ lattice with 
$am_\infty=0.5656$ (i.e., $a=0.16 (\lambda_*/2$), $L=64a=13.8 \lambda_\mu$)
and 
$\mathcal N_{\mathcal W} =320$ 
($=\mathcal N_Z\times \mathcal N_{\varphi}=20\times16$).
The initialization of the $\mathcal W$ fields is chosen with parameter
$\sigma
=0.1$ resulting in equal initial field amplitudes per gluon degree
of freedom for both gauge groups.
Shown are the total field energy densities as well as the contributions from transverse and longitudinal chromo-electric and chromo-magnetic fields.
In the lower panel, the individual runs are shown by light curves.
\label{fig:comparison1}}
\end{figure}

For initially small fields,
Figure~\ref{fig:comparison1} shows a comparison 
between SU(2) and
SU(3) simulations.
The simulations have been initialized with
random fluctuations in the $\mathcal W$ field with strength
$\sigma
=0.1$ for both gauge groups.
Depicted is the total energy density and its contributions
from chromo-electric and chromo-magnetic fields, each decomposed
into parts transverse and
longitudinal with respect to the direction of anisotropy. 
The SU(2) result has been obtained from averaging over 6 runs
with different initial conditions, while for SU(3) 3 runs have been used.
Because there are 8 gluons
in SU(3) compared to 3 gluons in SU(2) that are initialized with
the same average gauge field amplitude, the initial total
energy density of SU(3) is
larger than the SU(2) energy density by a factor 8/3.
Remarkably, at the end of the exponentially growing phase 
around $m_\infty \, t\sim60$
when the approximately linear growth regime
sets in, the total energy densities 
approximately agree (after averaging over individual runs---%
the saturation energy of individual runs varies substantially,
in particular for smaller lattice size). 
We have checked in comparison runs restricted to an Abelian U(1) subgroup 
that this agreement is not caused by the lattice spacing chosen
and that we are safely below the compactness bound.
The plot in linear scale shows that
the SU(3) result subsequently grows somewhat faster than the SU(2) result.

\begin{figure}
\begin{center}\includegraphics[%
  clip,
  scale=0.65]{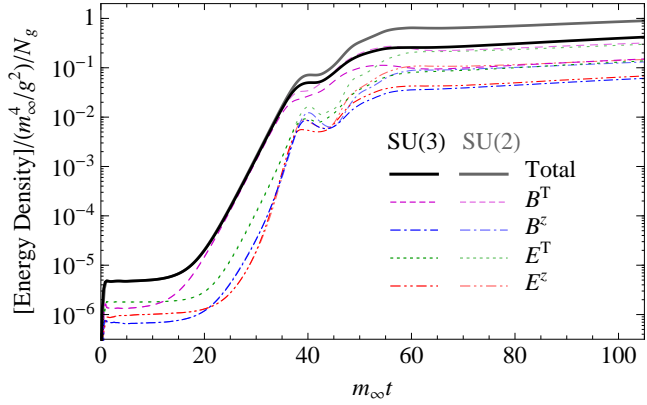} \includegraphics[%
  clip,
  scale=0.65]{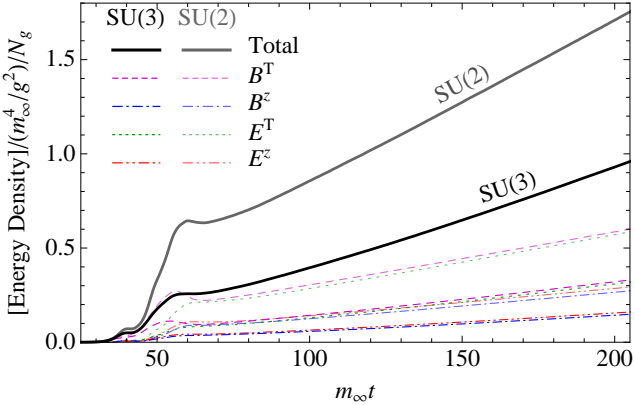}\end{center}
\caption{(Color online.) Same as Fig.~\ref{fig:comparison1} for the energy density divided
by the number of gluons $N_g$.\label{fig:comparison2}}
\end{figure}

In Fig.~\ref{fig:comparison2} the same results are compared by dividing by the number of gluons $N_g$. 
Now the SU(2) and SU(3) curves agree in the exponentially growing phase, with the SU(3) result starting to lag behind the SU(2) result 
from $m_\infty \, t\sim35$ onwards.
In the linear growth regime, it is now the SU(2) curve that grows faster. 
Therefore the dependence of the linear growth rate
in the non-Abelian regime on $N_c$
seems to be a factor 
between 1 and $N_g$.


\begin{figure}
\begin{center}\includegraphics[%
  clip,
  scale=0.65]{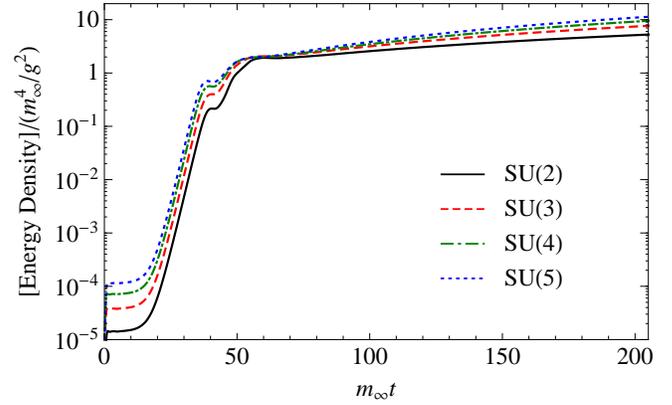} \includegraphics[%
  clip,
  scale=0.65]{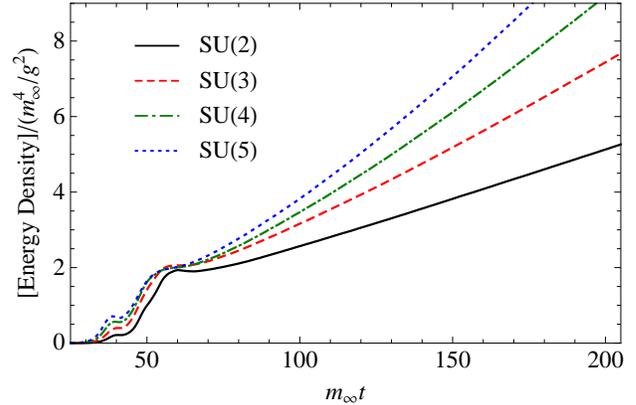}\end{center}
\caption{(Color online.) Comparison of average 
total field energy densities $\mathcal{E}$ for SU(2) through SU(5) on logarithmic (upper panel) and linear (lower panel) scale in
3+1 dimensional simulations for anisotropy parameter $\xi=10$ on
a $64^{3}$ lattice with 
$\mathcal N_{\mathcal W}=320$ 
($=N_Z\times N_{\phi}=20\times16$).
The initialization of the $\mathcal W$ fields is chosen with parameter
$\sigma_{\rm SU(2)}=\sigma_{\rm SU(3)}=\sigma_{\rm SU(4)}=\sigma_{\rm SU(5)}=0.1$.
\label{fig:energy1}}
\end{figure}

In order to study the systematics of the scaling with $N_c$
in the non-Abelian regime, we extend the gauge group SU($N_c$) to larger values of $N_c$. Figure \ref{fig:energy1} 
shows a comparison for gauge groups SU(2) through SU(5). This confirms
the observation of Fig.~\ref{fig:comparison1} that the
energy densities stop to grow at approximately the same value for 
different gauge groups,
and also that the energy densities of the larger gauge groups grow faster in the regime of linear growth.

\begin{figure}
\begin{center}
\includegraphics[%
  clip,
  scale=0.65]{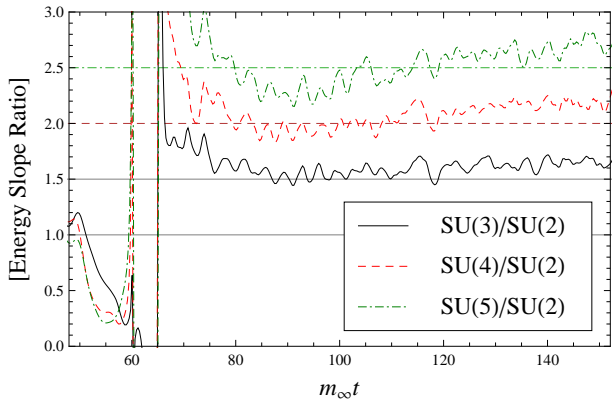}
\end{center}
\caption{(Color online.) The ratio of the time derivative of the energy densities of Fig.~\ref{fig:energy1} for gauge groups SU(3), SU(4), and SU(5), divided by the one of SU(2). For large times, this is seen to scale approximately
with $N_c$. Horizontal lines at 3/2, 4/2, and 5/2 mark the ratios for scaling by the number of colors.
\label{fig:energy3}}
\end{figure}

In Fig.~\ref{fig:energy3} we plot the ratio of the time derivatives of the energy densities of Fig.~\ref{fig:energy1} for the various gauge groups, all with same lattice parameters and initial gauge field strengths. 
Since the discrete derivatives wiggle strongly, the energy densities have been smoothed by averaging over a rectangular window function of size $m_\infty \Delta t=2$.
These results indicate that the slope of the energy density in the linear growth regime
scales approximately proportional to the number of colors, $N_c$. 

\begin{figure}[t]
\begin{center}\includegraphics[%
  clip,
  scale=0.65]{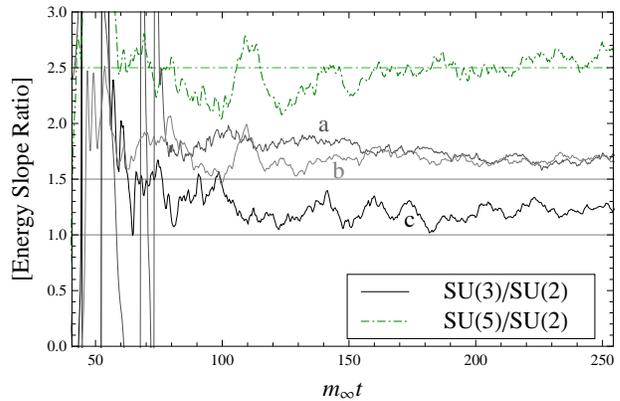}  \end{center}
\caption{(Color online.) The ratio of the time derivative of the energy densities for various lattice parameters.
(a) $32^3$, $am_\infty=0.2828$,
$\sigma=0.0707$;
(b) $32^3$, $am_\infty=0.2828$,
$\sigma=3.46$;
(c) $64^3$, $am_\infty=0.1414$,
$\sigma=0.1$.
For the SU(5)/SU(2) slope ratio, the same parameters as for (b) have been used.
\label{fig:slopes}}
\end{figure}

In Fig.~\ref{fig:slopes} we show
the slope ratio of gauge groups SU(3) and SU(2) for finer lattices (higher
ultraviolet cutoff) at equal (but now smaller) physical volume size.
For these curves, a rectangular window function of size $m_\infty \Delta t=10$ has been used.
The data summarize averages over (a) 20, (b) 8, (c) 3 runs
with different random initial conditions for SU(2), (a) 3,
(b) 4, (c) 1 runs for SU(3), and one run for SU(5).
In these runs, which have less resolution in the infrared,
we still find ratios that are consistently 
larger than 1, but more scattered about
the ratio of the number of colors.
A scaling by number of colors is also supported by the
slope ratio of SU(5) and SU(2) in Fig.~\ref{fig:slopes}.
We are therefore led to
the conjecture that the linear growth rate in the non-Abelian
regime is proportional to $N_c$, but have to
concede that our numerical verification involves sizeable
uncertainties.

\subsection{Spectra}

In order to get a glimpse of the underlying dynamics,
following Ref.~\cite{Arnold:2005ef} we consider
distribution functions of modes
\begin{eqnarray}
f_A(k) &=& \frac{k}{N_{\rm dof}V} \langle \mathbf{A}^2(k) \rangle\nonumber,\\ 
f_E(k) &=& \frac{1}{N_{\rm dof}kV} \langle \mathbf{E}^2(k) \rangle, \label{eq:distribution}
\end{eqnarray}
where $V$ is the total spatial volume and $N_{\rm dof}=2 N_g$ accounts for two transverse
polarization states and $N_g=N_c^2-1$ adjoint color states of the SU($N_c$) gauge theory.
The spectra are obtained from a Fourier transformation
of the $\mathbf A$ and $\mathbf E$ fields in the lattice Coulomb gauge \cite{Moore:1997cr},
which fixes the residual gauge freedom of the temporal axial gauge by minimizing unphysical high-momentum noise within 3-dimensional time
slices.
This turns out to be essential for our results for the spectra which would
otherwise show much more power in the ultraviolet.
Stable plasma modes are expected to contribute equally to
$f_A$ and $f_E$, whereas unstable modes are predominantly
magnetic, leading to $f_A>f_E$ in the corresponding momentum range.

\begin{figure}
\begin{center}\includegraphics[%
  clip,
  scale=0.65]{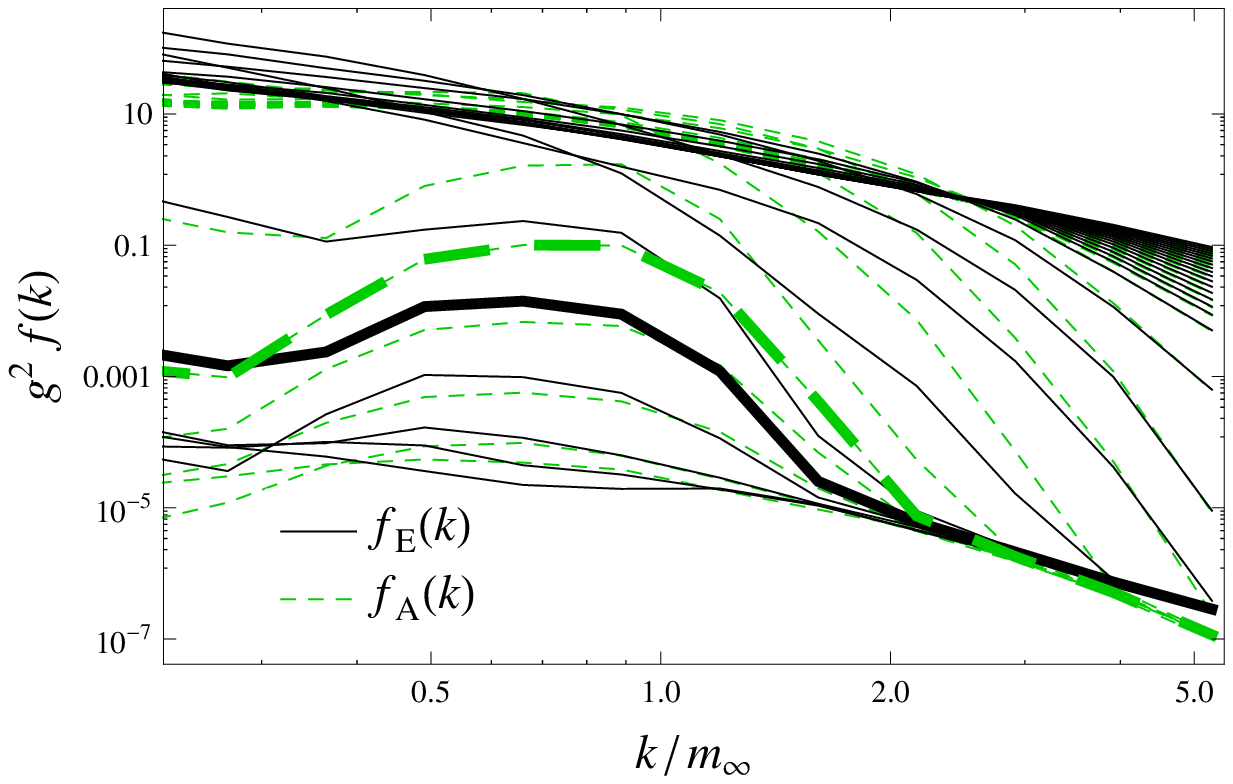} \includegraphics[%
  clip,
  scale=0.65]{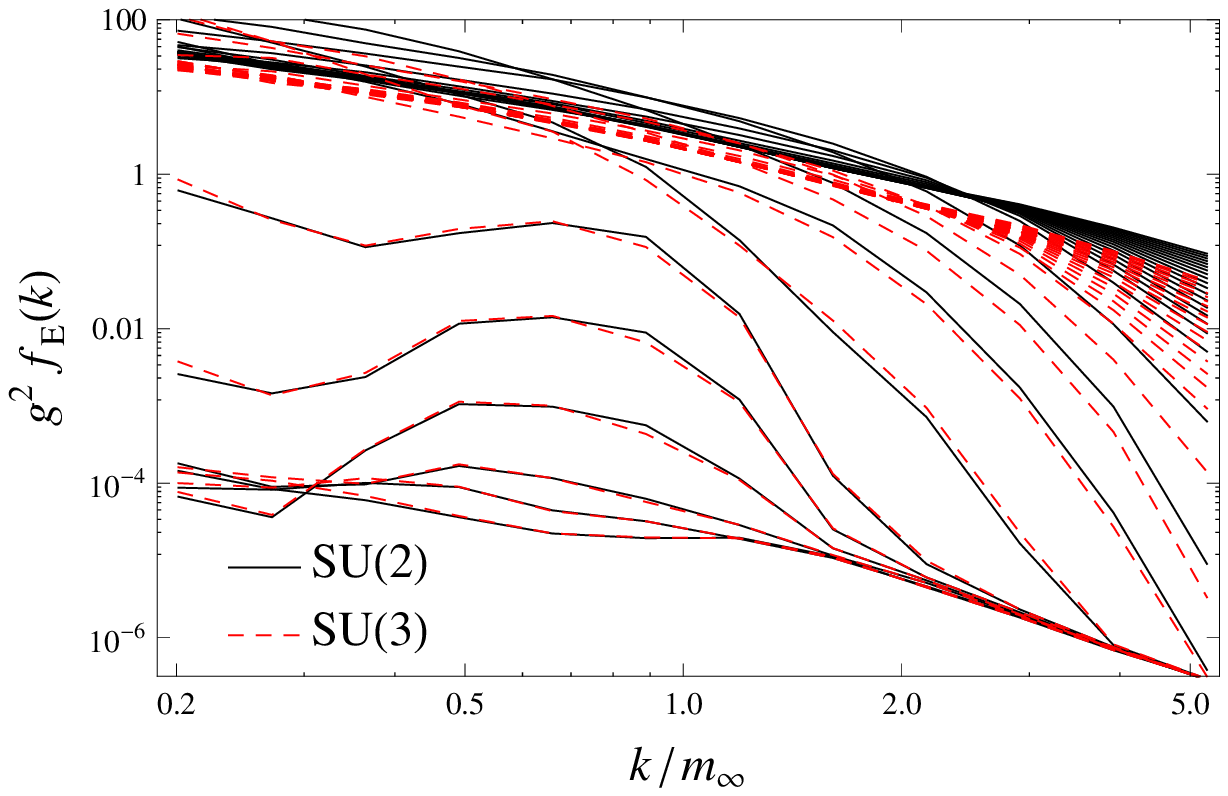} \end{center}
\caption{(Color online.) Spectra corresponding to Fig.~\ref{fig:comparison1}
for times $0 \leq m_\infty t \lesssim 150$.
The distance between the lines is $m_\infty \Delta t \approx 6$.
The upper panel shows the SU(2) spectra for the
magnetic ($f_A(k)$) and electric ($f_E(k)$) distribution functions.
The curves at time $m_\infty t \approx 23$ are plotted with thick lines.
The lower panel shows the electric ($f_E(k)$) distribution functions
for SU(2) vs.~SU(3).
\label{fig:spectra1}}
\end{figure}

Figure~\ref{fig:spectra1} shows the corresponding spectra,
normalized per gluonic degree of freedom. 
The results are averaged over lattice vectors $\mathbf k$ in
12 equally sized bins 
on a logarithmic scale along $|\mathbf k^2|$. In addition,
SU(2) curves represent averages over 6 runs,
while SU(3) curves represent averages over 4 runs.

The upper panel of Fig.~\ref{fig:spectra1} compares the 
magnetic ($f_A(k)$) and electric ($f_E(k)$) distribution functions.
The lowest curves correspond to the initial time.
Around
$k/m_\infty\sim 0.9$
there is a conspicuous exponential growth of modes 
associated with Weibel instabilities.
In order to ease the comparison of $f_A$ and
$f_E$ results, 
the time $m_\infty \Delta t \approx 14$ is singled out 
in this plot by thick lines. As can be seen also in 
Fig.~\ref{fig:comparison1}, the magnetic contributions dominate
over electric contributions
during the exponential growth phase.
When non-Abelian self-interaction of the gluonic fields sets in,
the exponential growth ceases, and only the higher modes
$k/m_\infty \gtrsim 2$ continue to grow, albeit more slowly. 

A comparison between SU(2) and SU(3) is shown in the lower panel of
Fig.~\ref{fig:spectra1}.
As expected, the behavior of 
the exponentially growing modes agrees well between the SU(2) and SU(3)
calculations since in that regime non-Abelian self interactions
are for the most part small. 
Note that the definitions (\ref{eq:distribution}) for $f_A(k)$ and $f_E(k)$ include a scaling by the number of degrees of freedom,
which is proportional to $N_g$.
This plot therefore corresponds to the energy density divided by $N_g$ as depicted in Fig.~\ref{fig:comparison2}.
The evolution of the SU(2) and SU(3) curves only deviate from each other
when non-linearities set in.


\begin{figure}
\begin{center}\includegraphics[%
  clip,
  scale=0.65]{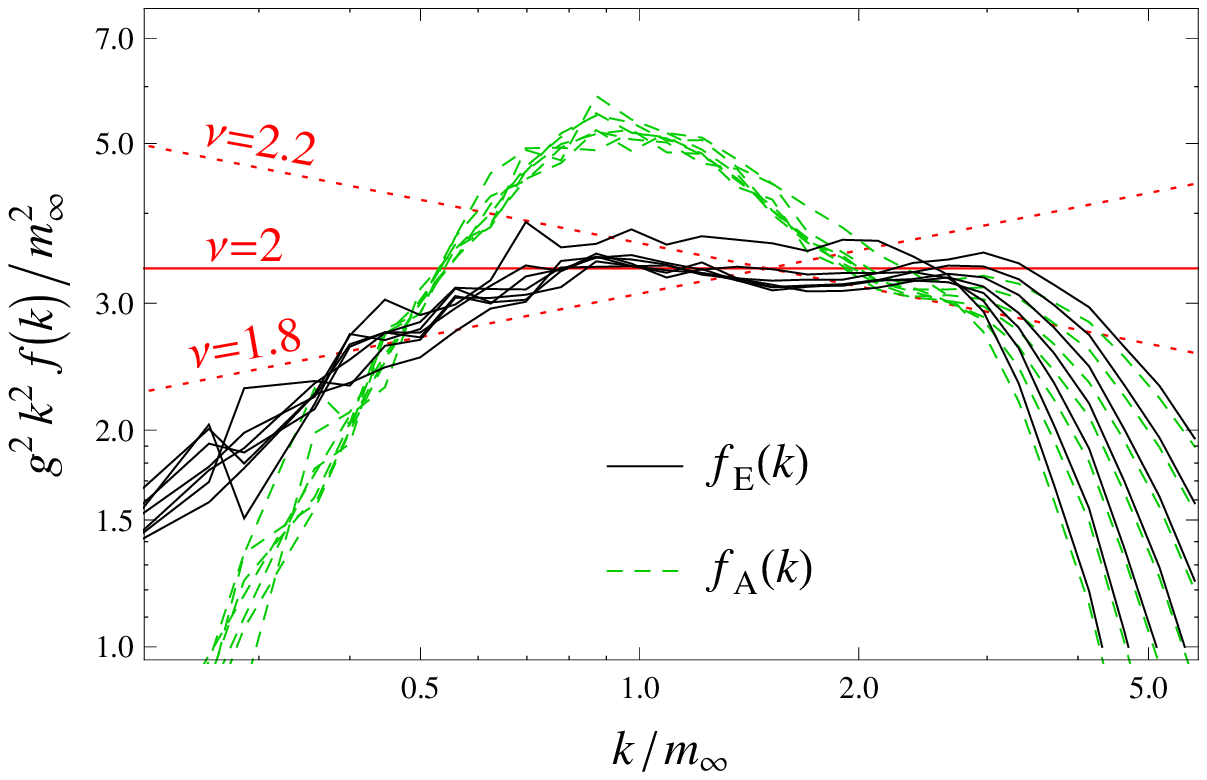} \includegraphics[%
  clip,
  scale=0.65]{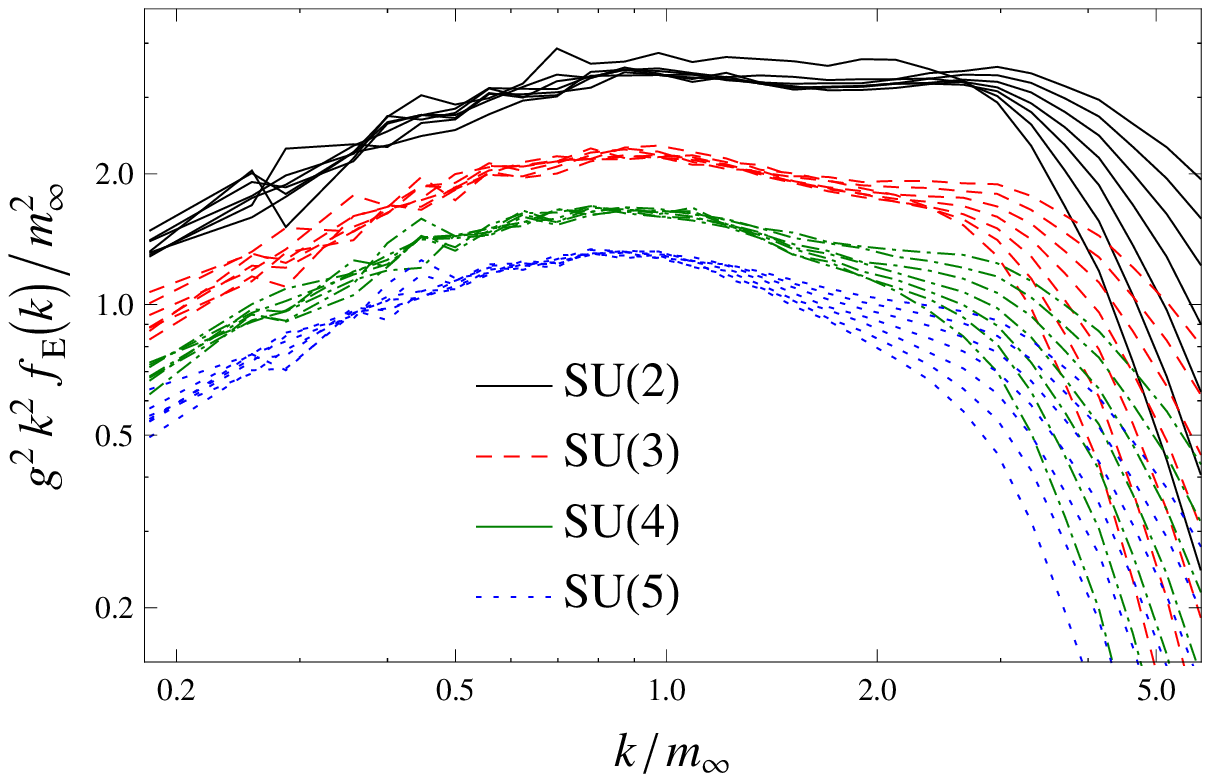}\end{center}
\caption{(Color online.) The power spectrum for SU(2) of the Coulomb gauge distribution $f(k)$
at late times $80 \lesssim m_\infty t \lesssim 150$.
The distance between the lines is $m_\infty \Delta t \approx 11$.
The upper panel shows a comparison for SU(2) for $f_E(k)$ and $f_A(k)$.
The straight central horizontal red line indicates a power law spectrum $f\sim k^{-\nu}$ with 
$\nu=2$, while the dotted lines correspond to $\nu=1.8$ and 2.2.
The lower panel compares the spectra $f_E(k)$ for SU(2), SU(3), SU(4), and SU(5).
\label{fig:spectra2}}
\end{figure}

In order to study the non-Abelian regime
with approximately linear growth of energies 
more carefully, Fig.~\ref{fig:spectra2}
depicts the spectra multiplied by $k^2$ and
at late-time times $80 \lesssim m_\infty t \lesssim 150$.
The distribution functions have been averaged over 32 bins along $|\mathbf k^2|$ in these plots 
so that the large $k$ region is better resolved.
The slow growth at large momenta $k/m_\infty \gtrsim 2$ corresponds to the linear growth regime of 
Fig.~\ref{fig:comparison1}.
On larger scales $k/m_\infty \gtrsim 2$ we have $f_A(k)\simeq f_E(k)$ as expected from stable modes in the perturbative regime.
In an intermediate range $0.5 \lesssim k/m_\infty \lesssim 2$, magnetic fields dominate over electric ones $f_A(k) > f_E(k)$,
which is caused by the predominantly magnetic plasma instabilities.

The straight lines in the upper panel of Fig.~\ref{fig:spectra2} correspond to simple power laws
$f(k)\sim k^{-\nu}$ with $\nu=2$ (straight line) and $\nu=1.8$, and 2.2 (dotted lines).
A distinct
power law behavior $\sim k^{-\nu}$ with $\nu\approx 2$
over a large range of momenta can be most clearly
extracted
from the electric distribution function $f_E(k)$, which does not
show a hump caused by 
the unstable magnetic modes, and if one excludes the higher momentum
modes close to the lattice cutoff and also the largest times, where
effects from the lattice cutoff may be felt.
As error bar for $\nu\approx2$ we infer $1.8 \lesssim \nu \lesssim 2.4$.

The lower panel of Fig.~\ref{fig:spectra2} compares the electric field spectra $f_E(k)$ at various gauge groups SU(2) through SU(5). The spectra of the various gauge groups appear to 
be similar, apart from an overall scaling factor.

\section{Conclusions}

We have studied for the first time 3+1 dimensional simulations of
non-Abelian plasma instabilities in the hard-loop framework
for
the gauge group SU(3) that is physically relevant
for the physics of heavy-ion collisions.
In order to study the dependence on the number of colors, we have
also considered gauge groups SU(4) and SU(5).
We have found that small seed fields which correspond to
the same amount of initial energy density per gluon degree of freedom
lead to comparable total energy density when exponential
growth of plasma instabilities is stopped by nonabelian
self-interactions and a phase of approximately linear growth begins.
We have confirmed that in all nonabelian gauge groups considered
a power-law spectrum $f(k)\sim k^{-\nu}$ with $\nu\approx 2$
develops for the higher-momentum modes, corresponding to
a cascade of energy towards the ultraviolet. The growth rate
of energy densities in this phase was found to scale roughly
proportional to the number of colors.

\section*{Acknowledgments}

This work has been supported by the Austrian Science Foundation
FWF, project no.\ P19526.
We would like to thank Paul Romatschke for communications and collaboration
on the SU(2) code
that was adapted for this work, and Maximilian Attems for
assistance. We also gratefully acknowledge valuable discussions with Peter Arnold, J\"urgen Berges,
Dietrich B\"odeker, Guy Moore, Al Mueller, Kari Rummukainen, and
Larry Yaffe. The numerical
calculations have been performed on
the ECT{*} Teraflop cluster, at the MPI Heidelberg parallel cluster,
and at the Vienna Scientific Cluster.

\bibliographystyle{prsty}
\bibliography{su3}

\begin{thebibliography}{10}

\bibitem{Heinz:2004pj}
U.~W. Heinz, AIP Conf. Proc. {\bf 739},  163  (2005).

\bibitem{Weibel:1959zz}
E.~S. Weibel, Phys. Rev. Lett. {\bf 2},  83  (1959).

\bibitem{Romatschke:2003ms}
P. Romatschke and M. Strickland, Phys. Rev. {\bf D68},  036004  (2003).

\bibitem{Romatschke:2004jh}
P. Romatschke and M. Strickland, Phys. Rev. {\bf D70},  116006  (2004).

\bibitem{Califano:2001}
F. Califano, N. Attico, F. Pegoraro, G. Bertin, and S.~V. Bulanov, Phys. Rev.
  Lett. {\bf 86},  5293  (2001).


\bibitem{Mrowczynski:1988dz}
S. Mrowczynski, Phys. Lett. {\bf B214},  587  (1988).

\bibitem{Mrowczynski:1993qm}
S. Mrowczynski, Phys. Lett. {\bf B314},  118  (1993).

\bibitem{Mrowczynski:2000ed}
S. Mrowczynski and M.~H. Thoma, Phys. Rev. {\bf D62},  036011  (2000).

\bibitem{Randrup:2003cw}
J. Randrup and S. Mrowczynski, Phys. Rev. {\bf C68},  034909  (2003).

\bibitem{Pokrovsky:1988bm}
Y.~E. Pokrovsky and A.~V. Selikhov, JETP Lett. {\bf 47},  12  (1988).

\bibitem{Pokrovsky:1990sz}
Y.~E. Pokrovsky and A.~V. Selikhov, Sov. J. Nucl. Phys. {\bf 52},  146  (1990).

\bibitem{Pokrovsky:1990uh}
Y.~E. Pokrovsky and A.~V. Selikhov, Sov. J. Nucl. Phys. {\bf 52},  385  (1990).

\bibitem{Pavlenko:1991ih}
O.~P. Pavlenko, Sov. J. Nucl. Phys. {\bf 55},  1243  (1992).

\bibitem{Mrowczynski:1994xv}
S. Mrowczynski, Phys. Rev. {\bf C49},  2191  (1994).

\bibitem{Mrowczynski:1996vh}
S. Mrowczynski, Phys. Lett. {\bf B393},  26  (1997).

\bibitem{Arnold:2003rq}
P. Arnold, J. Lenaghan, and G.~D. Moore, JHEP {\bf 08},  002  (2003).

\bibitem{Arnold:2004ti}
P. Arnold, J. Lenaghan, G.~D. Moore, and L.~G. Yaffe, Phys. Rev. Lett. {\bf
  94},  072302  (2005).

\bibitem{Rebhan:2004ur}
A. Rebhan, P. Romatschke, and M. Strickland, Phys. Rev. Lett. {\bf 94},  102303
   (2005).

\bibitem{Arnold:2005vb}
P. Arnold, G.~D. Moore, and L.~G. Yaffe, Phys. Rev. {\bf D72},  054003  (2005).

\bibitem{Rebhan:2005re}
A. Rebhan, P. Romatschke, and M. Strickland, JHEP {\bf 09},  041  (2005).

\bibitem{Dumitru:2007rp}
A. Dumitru, Y. Nara, B. Schenke, and M. Strickland, Phys. Rev. {\bf C78},
  024909  (2008).

\bibitem{Dumitru:2006pz}
A. Dumitru, Y. Nara, and M. Strickland, Phys. Rev. {\bf D75},  025016  (2007).

\bibitem{Dumitru:2005gp}
A. Dumitru and Y. Nara, Phys. Lett. {\bf B621},  89  (2005).

\bibitem{Romatschke:2005pm}
P. Romatschke and R. Venugopalan, Phys. Rev. Lett. {\bf 96},  062302  (2006).

\bibitem{Romatschke:2006nk}
P. Romatschke and R. Venugopalan, Phys. Rev. {\bf D74},  045011  (2006).

\bibitem{Romatschke:2006wg}
P. Romatschke and A. Rebhan, Phys. Rev. Lett. {\bf 97},  252301  (2006).

\bibitem{Rebhan:2008uj}
A. Rebhan, M. Strickland, and M. Attems, Phys. Rev. {\bf D78},  045023  (2008).

\bibitem{Rebhan:2009ku}
A. Rebhan and D. Steineder, Phys. Rev. {\bf D81},  085044  (2010).

\bibitem{Bodeker:2007fw}
D. B{\"o}deker and K. Rummukainen, JHEP {\bf 07},  022  (2007).

\bibitem{Arnold:2007cg}
P. Arnold and G.~D. Moore, Phys. Rev. {\bf D76},  045009  (2007).

\bibitem{Arnold:2005ef}
P.~B. Arnold and G.~D. Moore, Phys. Rev. {\bf D73},  025006  (2006).

\bibitem{Berges:2007re}
J. Berges, S. Scheffler, and D. Sexty, Phys. Rev. {\bf D77},  034504  (2008).

\bibitem{Berges:2008mr}
J. Berges, S. Scheffler, and D. Sexty, Phys. Lett. {\bf B681},  362  (2009).

\bibitem{Arnold:2005qs}
P.~B. Arnold and G.~D. Moore, Phys. Rev. {\bf D73},  025013  (2006).

\bibitem{Mueller:private}
A.~H. Mueller, private communication.


\bibitem{Bodeker:2005nv}
D. B{\"o}deker, JHEP {\bf 10},  092  (2005).

\bibitem{Mueller:2005un}
A.~H. Mueller, A.~I. Shoshi, and S.~M.~H. Wong, Phys. Lett. {\bf B632},  257
  (2006).

\bibitem{Mueller:2005hj}
A.~H. Mueller, A.~I. Shoshi, and S.~M.~H. Wong, Eur. Phys. J. {\bf A29},  49
  (2006).

\bibitem{Asakawa:2006tc}
M. Asakawa, S.~A. Bass, and B. Muller, Phys. Rev. Lett. {\bf 96},  252301
  (2006).

\bibitem{Mueller:2006up}
A.~H. Mueller, A.~I. Shoshi, and S.~M.~H. Wong, Nucl. Phys. {\bf B760},  145
  (2007).

\bibitem{Majumder:2006wi}
A. Majumder, B. Muller, and S.~A. Bass, Phys. Rev. Lett. {\bf 99},  042301
  (2007).

\bibitem{Carrington:2010sz}
M.~E. Carrington and A. Rebhan, arXiv:1011.0393, 2010.


\bibitem{Berges:2008zt}
J. Berges, D. Gelfand, S. Scheffler, and D. Sexty, Phys. Lett. {\bf B677},  210
   (2009).

\bibitem{Blaizot:2001nr}
J.-P. Blaizot and E. Iancu, Phys. Rept. {\bf 359},  355  (2002).

\bibitem{Pisarski:1997cp}
R.~D. Pisarski, hep-ph/9710370, 1997.


\bibitem{Mrowczynski:2004kv}
S. Mrowczynski, A. Rebhan, and M. Strickland, Phys. Rev. {\bf D70},  025004
  (2004).

\bibitem{Blaizot:1993be}
J.~P. Blaizot and E. Iancu, Nucl. Phys. {\bf B417},  608  (1994).

\bibitem{Moore:1997cr}
G.~D. Moore and N. Turok, Phys.Rev. {\bf D56},  6533  (1997).


\end{thebibliography}

\end{document}